\documentclass[aps,prd,twocolumn,showpacs,superscriptaddress,groupedaddress,amsmath,amssymb]{revtex4}
\usepackage{graphicx}
\usepackage{dcolumn}
\usepackage{bm}
\usepackage{amssymb}
\usepackage{multirow}

\newcommand {\Bd} {\ensuremath{B^0_d}}
\newcommand {\Bs} {\ensuremath{B^0_s}}
\newcommand {\Bq} {\ensuremath{B^0_q}}
\newcommand {\Ds} {\ensuremath{D_s}}

\newcommand {\barBq} {\ensuremath{\bar{B}^0_q}}
\newcommand {\asld} {\ensuremath{a^d_{\mathrm{sl}}}}
\newcommand {\asls} {\ensuremath{a^s_{\mathrm{sl}}}}
\newcommand {\aslq} {\ensuremath{a^q_{\mathrm{sl}}}}
\newcommand {\aslb} {\ensuremath{A^b_{\mathrm{sl}}}}
\newcommand {\ktomu} {\ensuremath{K \to \mu}}

\begin{document}

\widetext

\hspace{5.2in} \mbox{Fermilab-Pub-10/217-E}

\title{Evidence for an anomalous like-sign dimuon charge asymmetry}
\affiliation{Universidad de Buenos Aires, Buenos Aires, Argentina}
\affiliation{LAFEX, Centro Brasileiro de Pesquisas F{\'\i}sicas, Rio de Janeiro, Brazil}
\affiliation{Universidade do Estado do Rio de Janeiro, Rio de Janeiro, Brazil}
\affiliation{Universidade Federal do ABC, Santo Andr\'e, Brazil}
\affiliation{Instituto de F\'{\i}sica Te\'orica, Universidade Estadual Paulista, S\~ao Paulo, Brazil}
\affiliation{Simon Fraser University, Vancouver, British Columbia, and York University, Toronto, Ontario, Canada}
\affiliation{University of Science and Technology of China, Hefei, People's Republic of China}
\affiliation{Universidad de los Andes, Bogot\'{a}, Colombia}
\affiliation{Charles University, Faculty of Mathematics and Physics, Center for Particle Physics, Prague, Czech Republic}
\affiliation{Czech Technical University in Prague, Prague, Czech Republic}
\affiliation{Center for Particle Physics, Institute of Physics, Academy of Sciences of the Czech Republic, Prague, Czech Republic}
\affiliation{Universidad San Francisco de Quito, Quito, Ecuador}
\affiliation{LPC, Universit\'e Blaise Pascal, CNRS/IN2P3, Clermont, France}
\affiliation{LPSC, Universit\'e Joseph Fourier Grenoble 1, CNRS/IN2P3, Institut National Polytechnique de Grenoble, Grenoble, France}
\affiliation{CPPM, Aix-Marseille Universit\'e, CNRS/IN2P3, Marseille, France}
\affiliation{LAL, Universit\'e Paris-Sud, CNRS/IN2P3, Orsay, France}
\affiliation{LPNHE, Universit\'es Paris VI and VII, CNRS/IN2P3, Paris, France}
\affiliation{CEA, Irfu, SPP, Saclay, France}
\affiliation{IPHC, Universit\'e de Strasbourg, CNRS/IN2P3, Strasbourg, France}
\affiliation{IPNL, Universit\'e Lyon 1, CNRS/IN2P3, Villeurbanne, France and Universit\'e de Lyon, Lyon, France}
\affiliation{III. Physikalisches Institut A, RWTH Aachen University, Aachen, Germany}
\affiliation{Physikalisches Institut, Universit{\"a}t Freiburg, Freiburg, Germany}
\affiliation{II. Physikalisches Institut, Georg-August-Universit{\"a}t G\"ottingen, G\"ottingen, Germany}
\affiliation{Institut f{\"u}r Physik, Universit{\"a}t Mainz, Mainz, Germany}
\affiliation{Ludwig-Maximilians-Universit{\"a}t M{\"u}nchen, M{\"u}nchen, Germany}
\affiliation{Fachbereich Physik, Bergische  Universit{\"a}t Wuppertal, Wuppertal, Germany}
\affiliation{Panjab University, Chandigarh, India}
\affiliation{Delhi University, Delhi, India}
\affiliation{Tata Institute of Fundamental Research, Mumbai, India}
\affiliation{University College Dublin, Dublin, Ireland}
\affiliation{Korea Detector Laboratory, Korea University, Seoul, Korea}
\affiliation{SungKyunKwan University, Suwon, Korea}
\affiliation{CINVESTAV, Mexico City, Mexico}
\affiliation{FOM-Institute NIKHEF and University of Amsterdam/NIKHEF, Amsterdam, The Netherlands}
\affiliation{Radboud University Nijmegen/NIKHEF, Nijmegen, The Netherlands}
\affiliation{Joint Institute for Nuclear Research, Dubna, Russia}
\affiliation{Institute for Theoretical and Experimental Physics, Moscow, Russia}
\affiliation{Moscow State University, Moscow, Russia}
\affiliation{Institute for High Energy Physics, Protvino, Russia}
\affiliation{Petersburg Nuclear Physics Institute, St. Petersburg, Russia}
\affiliation{Stockholm University, Stockholm and Uppsala University, Uppsala, Sweden }
\affiliation{Lancaster University, Lancaster LA1 4YB, United Kingdom}
\affiliation{Imperial College London, London SW7 2AZ, United Kingdom}
\affiliation{The University of Manchester, Manchester M13 9PL, United Kingdom}
\affiliation{University of Arizona, Tucson, Arizona 85721, USA}
\affiliation{University of California Riverside, Riverside, California 92521, USA}
\affiliation{Florida State University, Tallahassee, Florida 32306, USA}
\affiliation{Fermi National Accelerator Laboratory, Batavia, Illinois 60510, USA}
\affiliation{University of Illinois at Chicago, Chicago, Illinois 60607, USA}
\affiliation{Northern Illinois University, DeKalb, Illinois 60115, USA}
\affiliation{Northwestern University, Evanston, Illinois 60208, USA}
\affiliation{Indiana University, Bloomington, Indiana 47405, USA}
\affiliation{Purdue University Calumet, Hammond, Indiana 46323, USA}
\affiliation{University of Notre Dame, Notre Dame, Indiana 46556, USA}
\affiliation{Iowa State University, Ames, Iowa 50011, USA}
\affiliation{University of Kansas, Lawrence, Kansas 66045, USA}
\affiliation{Kansas State University, Manhattan, Kansas 66506, USA}
\affiliation{Louisiana Tech University, Ruston, Louisiana 71272, USA}
\affiliation{University of Maryland, College Park, Maryland 20742, USA}
\affiliation{Boston University, Boston, Massachusetts 02215, USA}
\affiliation{Northeastern University, Boston, Massachusetts 02115, USA}
\affiliation{University of Michigan, Ann Arbor, Michigan 48109, USA}
\affiliation{Michigan State University, East Lansing, Michigan 48824, USA}
\affiliation{University of Mississippi, University, Mississippi 38677, USA}
\affiliation{University of Nebraska, Lincoln, Nebraska 68588, USA}
\affiliation{Rutgers University, Piscataway, New Jersey 08855, USA}
\affiliation{Princeton University, Princeton, New Jersey 08544, USA}
\affiliation{State University of New York, Buffalo, New York 14260, USA}
\affiliation{Columbia University, New York, New York 10027, USA}
\affiliation{University of Rochester, Rochester, New York 14627, USA}
\affiliation{State University of New York, Stony Brook, New York 11794, USA}
\affiliation{Brookhaven National Laboratory, Upton, New York 11973, USA}
\affiliation{Langston University, Langston, Oklahoma 73050, USA}
\affiliation{University of Oklahoma, Norman, Oklahoma 73019, USA}
\affiliation{Oklahoma State University, Stillwater, Oklahoma 74078, USA}
\affiliation{Brown University, Providence, Rhode Island 02912, USA}
\affiliation{University of Texas, Arlington, Texas 76019, USA}
\affiliation{Southern Methodist University, Dallas, Texas 75275, USA}
\affiliation{Rice University, Houston, Texas 77005, USA}
\affiliation{University of Virginia, Charlottesville, Virginia 22901, USA}
\affiliation{University of Washington, Seattle, Washington 98195, USA}
\author{V.M.~Abazov} \affiliation{Joint Institute for Nuclear Research, Dubna, Russia}
\author{B.~Abbott} \affiliation{University of Oklahoma, Norman, Oklahoma 73019, USA}
\author{M.~Abolins} \affiliation{Michigan State University, East Lansing, Michigan 48824, USA}
\author{B.S.~Acharya} \affiliation{Tata Institute of Fundamental Research, Mumbai, India}
\author{M.~Adams} \affiliation{University of Illinois at Chicago, Chicago, Illinois 60607, USA}
\author{T.~Adams} \affiliation{Florida State University, Tallahassee, Florida 32306, USA}
\author{E.~Aguilo} \affiliation{Simon Fraser University, Vancouver, British Columbia, and York University, Toronto, Ontario, Canada}
\author{G.D.~Alexeev} \affiliation{Joint Institute for Nuclear Research, Dubna, Russia}
\author{G.~Alkhazov} \affiliation{Petersburg Nuclear Physics Institute, St. Petersburg, Russia}
\author{A.~Alton$^{a}$} \affiliation{University of Michigan, Ann Arbor, Michigan 48109, USA}
% ...
%
%
\collaboration{The D0 Collaboration\footnote{with visitors from
$^{a}$  
Augustana College, Sioux Falls, SD, USA,
$^{b}$  
The University of Liverpool, Liverpool, UK,
$^{c}$  
SLAC, Menlo Park, CA, USA,
$^{d}$  
ICREA/IFAE, Barcelona, Spain,
$^{e}$  
Centro de Investigacion en Computacion - IPN,
  Mexico City, Mexico,
$^{f}$  
ECFM, Universidad Autonoma de Sinaloa, Culiac\'an, Mexico,
$^{g}$  
and Universit{\"a}t Bern, Bern, Switzerland.
}} \noaffiliation
\vskip 0.25cm
\date{\today}

\begin{abstract}
We measure the charge asymmetry $A\equiv(N^{++} - N^{--})/(N^{++} + N^{--})$ 
of like-sign dimuon events in 6.1~fb$^{-1}$
of $p\bar{p}$ collisions recorded with the D0 detector at a center-of-mass
energy $\sqrt{s}=1.96$~TeV at the Fermilab Tevatron collider. From $A$ we
extract the like-sign dimuon charge asymmetry in semileptonic $b$-hadron
decays: $\aslb=-0.00957\pm0.00251\thinspace({\rm stat})\pm0.00146\thinspace({\rm sys})$.
It differs by 3.2~standard deviations from the standard model prediction
$\aslb({\rm SM})=(-2.3^{+0.5}_{-0.6})\times10^{-4}$, and provides first
evidence of anomalous $CP$ violation in the mixing of neutral $B$ mesons.
\end{abstract}

\pacs{13.25.Hw; 14.40.Nd}
\maketitle

Studies of particle production and decay under the reversal of
discrete symmetries (charge, parity and time) have
yielded considerable insight into the structure of theories
that describe high energy phenomena. Of particular interest
is the observation of $CP$ violation, a phenomenon well
established in the $K^0$ and $B^0_d$ systems, but not
in the $B^0_s$ system where the effects of $CP$-violation
are expected to be small in the standard model
(SM)~\cite{Nierste}. A review of the experimental results and
of the theoretical framework for describing $CP$ violation in 
neutral mesons decays can be found in Ref.~\cite{pdg}.
The violation of $CP$ symmetry is a necessary
condition for baryogenesis, the process thought to
be responsible for the matter-antimatter asymmetry of the
universe~\cite{sakharov}. However, the observed level of $CP$ violation
in the $K^0$ and $B^0_d$ systems is not sufficient to accommodate
this asymmetry, suggesting the presence of additional sources
of $CP$ violation beyond the SM~\cite{gavela_huet}.

This Letter and a more detailed Article~\cite{d0asym2mu} present a measurement 
of the charge asymmetry for like-sign muon pairs.  The data, corresponding 
to an integrated luminosity of $6.1$ fb$^{-1}$, were recorded 
with the D0 detector~\cite{d0det} at the Fermilab Tevatron 
proton-antiproton ($p\bar{p}$) collider, operating at a center-of-mass 
energy of 1.96~TeV. The D0 experiment is well suited to the investigation of the
small effects of $CP$ violation because the periodic reversal
of the D0 solenoid and toroid magnetic field polarities results in a 
cancellation of most detector-related asymmetries. In addition, the 
$p\bar p$ initial state is a $CP$ eigenstate, and the high center-of-mass 
energy provides access to mass states beyond the reach of the 
$B$-factories running at $\sqrt{s}=M(\Upsilon(4S))$.

The like-sign dimuon charge asymmetry $A$ is defined as
\begin{equation}
A \equiv \frac{N^{++} - N^{--}}{N^{++} + N^{--}},
\label{o_defA}
\end{equation}
where $N^{++}$ and $N^{--}$ represent
the number of events in which the two muons of
highest transverse momentum, satisfying the kinematic selections
described below,
have the same positive or negative charges. After removing
contributions from background and from remaining detector effects, 
any residual asymmetry is assumed to arise solely from the mixing of 
$\Bq$ $(q=d,s)$ mesons (via $\Bq\leftrightarrow\barBq$ oscillations)
that later decay semileptonically. This corrected
asymmetry $\aslb$ is defined as
\begin{equation}
\aslb \equiv \frac{N^{++}_{b} - N^{--}_{b}}{N^{++}_{b} + N^{--}_{b}},
\end{equation}
where $N^{++}_{b}$ and $N^{--}_{b}$ represent the number of events
containing two $b$-hadrons decaying semileptonically into
two positive or two negative muons, respectively. Assuming $CPT$ invariance, 
each neutral $\Bq$ meson contributes a term to this asymmetry
\begin{eqnarray}
\aslb & = & \beta_d \asld + \beta_s \asls, \label{Ab_7} \\
\mathrm{with}\thickspace\aslq & = & \frac{\Delta \Gamma_q}{\Delta M_q} \tan \phi_q, \label{i_phiq}
\end{eqnarray}
where $\phi_q$ is the $CP$-violating phase, and $\Delta M_q$ and 
$\Delta \Gamma_q$ are the mass and width differences between the 
eigenstates of the propagation matrices of the neutral $\Bq$ mesons.
The values of $\beta_d = 0.506 \pm 0.043$ and $\beta_s = 0.494 \pm 0.043$ 
are taken from previous measurements~\cite{pdg}. The 
SM predicts $\phi_s = 0.0042 \pm 0.0014$, $\phi_d = -0.091^{+0.026}_{-0.038}$, 
$\Delta\Gamma_d = (26.7^{+5.8}_{-6.5})\times 10^{-4}\thinspace \mathrm{ps}^{-1}$,
$\Delta\Gamma_s = (9.6 \pm 3.9)\times 10^{-2}\thinspace \mathrm{ps}^{-1}$,
$\asld = (-4.8 ^{+1.0}_{-1.2}) \times 10^{-4}$ and
$\asls = (2.06 \pm 0.57) \times 10^{-5}$~\cite{Nierste}. Deviations
of the expected asymmetries from zero are negligible compared to the present experimental 
sensitivity, and correspond to a small value for $\aslb$~\cite{Nierste}:
\begin{equation}
\aslb({\rm SM}) = (-2.3^{+0.5}_{-0.6}) \times 10^{-4}.
\label{in_aslbsm}
\end{equation}
Extensions of the SM containing additional contributions to the 
Feynman ``box" diagrams responsible for $\Bq$ mixing
can result in larger values of $\phi_q$ or $\Delta\Gamma_q$~\cite{Randall,Hewett,Hou,Soni,Buras}. 
Measurements of $\aslb$, $\phi_q$ or $\Delta\Gamma_q$ that differ significantly from the
expectations of the SM would therefore be indicative of the presence of physics beyond the SM.

At the Fermilab Tevatron collider, $b$ quarks are produced mainly in 
$b \bar{b}$ pairs. In like-sign dimuon events, one muon arises from
direct semileptonic decay, e.g., $b \rightarrow \mu^- X$,  of a
$\barBq$ or $B^-$ meson, and the other muon from a $\Bq\leftrightarrow\barBq$ oscillation followed,
in this example, by a semileptonic decay of the $\barBq$ meson, 
$B^0_q \rightarrow \bar{B}^0_q \rightarrow \mu^- X$.
The main background for these measurements arises from events with at least 
one muon from kaon or pion decays or from the sequential decay of $b$ quarks,
$b \rightarrow c \rightarrow \mu^+ X$. The most important background
asymmetry arises from the fact that $K^+$ and $K^-$ mesons interact
differently with the material of the detector, and thus their
decay rates into positive and negative muons are not identical.

The asymmetry $\aslb$ can also be obtained from the measurement of
the charge asymmetry $a^b_{\rm sl}$ in 
semileptonic decays of $b$ quarks to muons of ``wrong charge",
i.e., a muon of charge opposite to the sign of the charge of the original $b$ quark, 
induced through $\Bq\leftrightarrow\barBq$ oscillations~\cite{Grossman}:
\begin{equation}
a^b_{\rm sl} \equiv \frac{\Gamma(\bar B \to B \to \mu^+ X) - \Gamma(B \to \bar B \to \mu^- X)}
             {\Gamma(\bar B \to B \to \mu^+ X) + \Gamma(B \to \bar B \to \mu^- X)} = \aslb.
\label{asl-w}
\end{equation}
The asymmetry $a^b_{\rm sl}$ can be measured from the
inclusive muon charge asymmetry
 \begin{equation}
a \equiv \frac{n^+ - n^-}{n^+ + n^-},
\label{asym_a}
\end{equation}
where $n^+$ and $n^-$ correspond to the number of positive 
and negative muons satisfying the kinematic selections. For the asymmetry 
$a$, the signal comes from $\Bq$ mixing, followed by the semileptonic decay. In addition
to the background already considered for $A$, the direct production
of $c$ quark pairs followed by their semileptonic decays constitutes an
additional source of muons contributing to~$a$. 

We define all muons from weak decays of $b$ and $c$ quarks as signal, and 
use the branching fractions and momentum spectra of particles in the
decay chains that produce such muons to determine the dilution of
the $\aslb$ asymmetry in the observed asymmetry of the signal component.
The dilutions, defined as the coefficients which relate the
signal asymmetries to $\aslb$, are $0.070\pm0.006$ and $0.486\pm0.032$ for the inclusive
muon and for the like-sign dimuon signal asymmetries, respectively.
The difference in the dilution coefficients arises because the
presence of the second muon with the same charge acts as a flavor tag. 
Therefore, the asymmetry $A$ is far more sensitive to $\aslb$ than $a$.

We measure the asymmetries $A$ and $a$ in the like-sign dimuon
and inclusive muon data, respectively. These have different contributions
from background processes and from detector asymmetries, which
are measured directly in data as a function of the muon transverse momenta,
and are used to correct the measured asymmetries. After
applying all corrections, the only expected source of residual asymmetry in
both the inclusive muon and dimuon samples is the asymmetry $\aslb$.
Given the difference in sensitivity between $A$ and $a$ and the fact that the asymmetry $a$
is dominated by detector effects, we do not take
a weighted average of the two determinations of \aslb. Instead, we use the
measurement of $a$ to constrain the background contributions to $A$, 
thereby achieving a further reduction of the total uncertainty on $\aslb$. 
This is possible because the detector effects and their related 
systematic uncertainties largely cancel in the linear combination
of $A$ and $a$.

The inclusive muon and like-sign dimuon samples are obtained from
data collected with single and dimuon triggers, respectively. Charged particles
with transverse momentum in the range $ 1.5 < p_T < 25$~GeV 
and with pseudorapidity $|\eta| < 2.2$~\cite{rapidity} are considered as
muon candidates. The upper limit on $p_T$ is applied to suppress 
the contribution of muons from $W$ and $Z$ boson decays. To ensure 
that the muon candidate passes through the detector, including 
all three layers of the muon system, we require either $p_T > 4.2$~GeV 
or a longitudinal momentum component $|p_z| > 6.4$~GeV. Muon candidates
are selected by matching central tracks with a segment reconstructed in the
muon system and by applying tight quality requirements aimed at reducing
false matching and background from cosmic rays and beam halo.
The transverse impact parameter of the muon track relative to the reconstructed
$p\bar p$ interaction vertex must be smaller than $0.3$~cm, with the
longitudinal distance from the point of closest approach to this vertex smaller than $0.5$ cm.
Strict quality requirements are also applied to the tracks and 
to the reconstructed $p\bar p$ interaction vertex. The inclusive muon sample contains
all muons passing the selection 
requirements. If an event contains more than one muon, each muon
is included in the inclusive muon sample. The like-sign dimuon
sample contains all events with at least two muon candidates
with the same charge. These two muons are required to have an invariant
mass greater than 2.8~GeV to minimize the number of events in which both muons originate
from the same $b$ quark. 

Muons from decays of charged kaons and pions 
and from incomplete absorption of hadrons that penetrate the calorimeter and
reach the muon detectors (``punch-through"), as well as false matches of central
tracks to segments reconstructed in the outer muon detector, are considered
as detector backgrounds. We use data to measure the fraction of each source of background in both
the dimuon and inclusive muon samples, and the corresponding asymmetries.
Data are also used to determine the intrinsic charge-detection asymmetry of the D0 detector.
Since the interaction length of the $K^+$ meson is greater than that of 
the $K^-$ meson~\cite{pdg}, kaons provide a positive contribution
to the asymmetries $A$ and $a$. The asymmetries for other background sources (pions,
protons and falsely reconstructed tracks) are at least a factor of ten smaller.

\begin{figure}
\begin{center}
\includegraphics[width=0.50\textwidth]{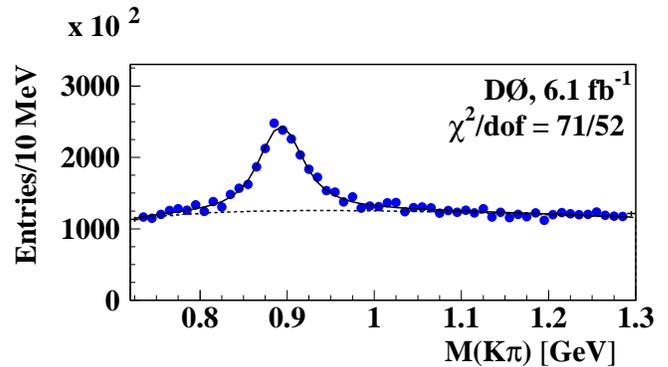}
\caption{The difference in the number of events for the $K^+ \pi^-$ and $K^- \pi^+$ mass distributions of $K^{*0}$
candidates in the inclusive muon sample. The solid line represents the result of the fit, while the dashed line
shows the background contribution.}
\label{fig_kst0_diff}
\end{center}
\end{figure}

The asymmetry for kaon tracks that are eventually misidentified as muons 
(\ktomu~tracks) is measured in data using $K^{*0} \to K^{\pm} \pi^{\mp}$ and $\phi \to K^+ K^-$ decays. 
For both channels we select muon candidates from the entire 
inclusive muon sample and examine mass distributions separately for
events with $K^+ \to \mu^+$ and $K^- \to \mu^-$ tracks, 
extracting the sum and the difference in the number of $K^{*0}$ or $\phi$ 
meson decays containing positive or negative \ktomu~ tracks.
The distribution of this difference as a function of the invariant
mass of the $K^{*0}$ candidates is shown in Fig.~\ref{fig_kst0_diff}.
The resulting asymmetry is corrected using simulations~\cite{d0asym2mu}
for the fraction of kaons ($\approx6\%$) that decay prior to being reconstructed. 
Similarly, the asymmetry for pion or proton tracks misidentified as muons are 
measured using samples of $K_S \to \pi^+ \pi^-$ and $\Lambda \to p \pi^-$ decays, 
respectively.

The fraction of muons from kaons is also determined from the $K^{*0} \to K^+ \pi^-$
sample. The fraction of all kaons arising from  $K^{*0}$ decay is
taken from the observed $K^{*\pm} \to K_S \pi^{\pm}$ decays using
the assumption of isospin invariance, which is validated in data~\cite{pdg}. 
The probability of identifying
the associated $\pi^{\pm}$ in the $K^{*0}$ decay is taken to be the
same as in $K^{*\pm}$, as is confirmed by simulation.
The fractions of pions and protons associated with identified muons,
relative to the fraction of muons from kaons, are estimated using 
the decays $K_S \to \pi^+ \pi^-$, $\phi \to K^+ K^-$, and $\Lambda \to p \pi^-$ 
decays, and the spectra and multiplicities of pions, kaons and protons from simulation.

After subtracting the muons originating from kaons, pions and protons, we find
the fraction of muons in the inclusive muon sample from prompt
sources constituting the signal sample (heavy flavor) to be
$0.581 \pm 0.014~({\rm stat}) \pm 0.039~({\rm syst})$.
The signal fraction arising from prompt sources for the
like-sign dimuon sample, after subtracting the contribution
from events where one or both muons are background, is
$0.665 \pm 0.016~({\rm stat}) \pm 0.033~({\rm syst})$.

Table~III of Ref.~\cite{D01} lists all significant contributions to the
dimuon charge asymmetry caused by detector effects. The largest
of these is $\approx3$\%. The reversal of both solenoid and toroid 
magnet polarities suppresses many detector effects, 
reducing thereby any charge asymmetry introduced by track reconstruction
and muon identification considerably~\cite{D01}. The small residual reconstruction
asymmetry is measured using a sample of $J/\psi \to \mu^+ \mu^-$ decays
reconstructed from two central detector tracks, with at least one matching a
track segment in the muon detector. This measurement is performed
as a function of the muon $p_T$ and indicates a residual detector asymmetry of order $10^{-3}$. 

The uncorrected asymmetries $a$ and $A$ are obtained by counting the number
of events of each charge in the inclusive muon and like-sign dimuon samples,
respectively. There are $1.495 \times 10^9$ muons in the inclusive
muon sample and $3.731 \times 10^6$ events in the like-sign dimuon sample.
The uncorrected asymmetries are
\begin{eqnarray}
\label{value_a}
a & = & +0.00955 \pm 0.00003~({\rm stat}), \\
A & = & +0.00564 \pm 0.00053~({\rm stat}).
\label{value_A}
\end{eqnarray}
After correcting for background and for the dilutions of
the $\aslb$ asymmetry in the observed asymmetries of
the signal component, we obtain
\begin{equation}
\aslb = +0.0094 \pm 0.0112~({\rm stat}) \pm 0.0214~({\rm syst})
\label{ah1} 
\end{equation}
from the inclusive muon sample and
\begin{equation}
\aslb = -0.00736 \pm 0.00266~({\rm stat}) \pm 0.00305~({\rm syst}) 
\label{ah2}
\end{equation}
from the like-sign dimuon sample. The uncertainties
on the measurement obtained from the like-sign dimuon sample
are much smaller, as expected from the factor of seven difference in the
dilution coefficients that relate the signal asymmetries to $\aslb$
in the two samples. Since the same background processes
contribute to the uncorrected asymmetries $a$ and $A$, their 
uncertainties are strongly correlated. We take advantage of this 
correlation to obtain a single optimized value of $\aslb$ with 
higher precision, by using a linear combination of the uncorrected 
asymmetries
\begin{equation}
A' \equiv A - \alpha a.
\label{combination}
\end{equation}
We scan the coefficient $\alpha$ in order to minimize the total
uncertainty on the value of $\aslb$, which occurs when $\alpha=0.959$.
The corresponding final result for the asymmetry $\aslb$ is:
\begin{equation}
\label{ah3}
\aslb = -0.00957 \pm 0.00251~({\rm stat}) \pm 0.00146~({\rm syst}).
\end{equation}
It differs by 3.2 standard deviations from the SM prediction for $\aslb$ of
Eq.~(\ref{in_aslbsm}). The contributions to the total uncertainty of 
$\aslb$ in Eqs.~(\ref{ah1}), (\ref{ah2}) and~(\ref{ah3}) are listed 
in Table~\ref{tab5}, and the result in Eq.~(\ref{ah3}) is dominated 
by statistical uncertainties.

\begin{table}
\caption{\label{tab5}
Sources of uncertainty on $\aslb$ in Eqs.~(\ref{ah1}),
(\ref{ah2}), and~(\ref{ah3}). The first eight rows correspond to statistical uncertainties 
and the next three rows to systematic uncertainties.}
\begin{ruledtabular}
\newcolumntype{A}{D{A}{\pm}{-1}}
\newcolumntype{B}{D{B}{-}{-1}}
\begin{tabular}{cccc}
Source & $\delta(\aslb)(\ref{ah1})$ & $\delta(\aslb)$(\ref{ah2}) & $\delta(\aslb)$(\ref{ah3}) \\
\hline
$A$ or $a$ (stat)       & 0.00066 & 0.00159 & 0.00179 \\
$K$ fraction (stat)     & 0.00222 & 0.00123 & 0.00140 \\
$\pi$ fraction          & 0.00234 & 0.00038 & 0.00010 \\
$p$ fraction            & 0.00301 & 0.00044 & 0.00011 \\
$K$ asymmetry           & 0.00410 & 0.00076 & 0.00061 \\
$\pi$ asymmetry         & 0.00699 & 0.00086 & 0.00035 \\
$p$ asymmetry           & 0.00478 & 0.00054 & 0.00001 \\
detector asymmetry      & 0.00405 & 0.00105 & 0.00077 \\
\hline
$K$ fraction (syst)     & 0.02137 & 0.00300 & 0.00128 \\
$\pi$, $K$, $p$
multiplicity            & 0.00098 & 0.00025 & 0.00018 \\
$\aslb$ dilution        & 0.00080 & 0.00046 & 0.00068 \\
\hline
Total statistical       & 0.01118 & 0.00266 & 0.00251 \\
Total systematic        & 0.02140 & 0.00305 & 0.00146 \\
Total                   & 0.02415 & 0.00405 & 0.00290
\end{tabular}
\end{ruledtabular}
\end{table}

\begin{figure}
\begin{center}
\includegraphics[width=0.50\textwidth]{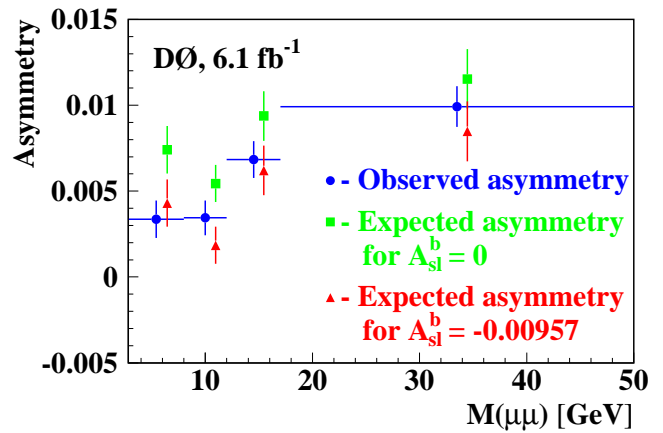}
\caption{ (Color online) The observed and expected like-sign dimuon charge asymmetry $A$
as a function of dimuon invariant mass. The expected asymmetry is shown
for $\mbox{\aslb=0}$ and $\mbox{\aslb=-0.00957}$.}
\label{a-mmm}
\end{center}
\end{figure}

Several consistency checks are performed by
dividing the data into smaller samples using additional selections 
based on data taking periods, muon and track quality requirements, and
changing the requirements on impact parameter, transverse momentum, 
polar angle and rapidity of the muons. The resulting variations of
$\aslb$ are statistically consistent with the result of Eq.~\ref{ah3},
even if the individual values of the uncorrected asymmetries 
$A$ and $a$ vary widely between the
different samples due to changes in the background contributions.
Both the size and the dependence on the muon momentum of
the asymmetry $a$, which is dominated by background, are
reproduced correctly through measurements of the background
fractions and asymmetries. Similarly, the dependence of the
like-sign dimuon asymmetry on the dimuon
invariant mass observed in data, as shown in Fig.~\ref{a-mmm}, 
is reproduced by expectations when $\aslb$ is
fixed to its measured value, while there are significant
discrepancies if $\aslb=0$ is assumed.

\begin{figure}
\begin{center}
\includegraphics[width=0.45\textwidth]{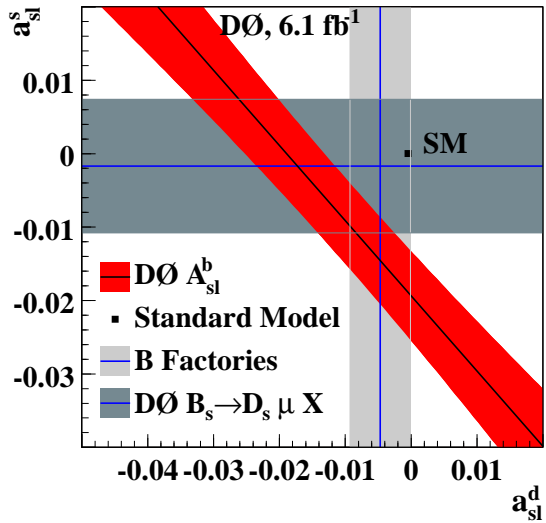}
\caption{(Color online) Comparison of $\aslb$ in data with the SM prediction
for $\asld$ and $\asls$. Also shown are other measurements of 
\mbox{$\asld=-0.0047\pm0.0046$}~\cite{babar,belle,hfag} and
\mbox{$\asls=-0.0017\pm0.0091$}~\cite{asl-d0}. The bands 
represent the $\pm 1$ standard deviation uncertainties on
each measurement.}
\label{asl_svsd}
\end{center}
\end{figure}

The measured value of $\aslb$ places
a constraint on the charge asymmetries in semileptonic
decays of $\Bd$ and $\Bs$ mesons and on the $CP$-violating phases
of the $\Bd$ and $\Bs$ mixing matrices, as given by Eqs.~\ref{Ab_7} and~\ref{i_phiq}.
Figure~\ref{asl_svsd} presents our measurement of $\aslb$ in the $\asld$--$\asls$ plane, together with
direct measurements of $\asld$ from the $B$-factories~\cite{babar,belle,hfag} and of
our independent measurement of $\asls$ in $\Bs \to \Ds \mu X$ decays~\cite{asl-d0}.
Additional comparisons and combinations of these results with previous
measurements sensitive to the same physics effect are given in Ref.~\cite{d0asym2mu}.

In conclusion, we have measured the like-sign dimuon charge asymmetry
$\aslb$ of semileptonic $b$-hadron decays:
\begin{equation}
\aslb=-0.00957\pm0.00251\thinspace({\rm stat})\pm0.00146\thinspace({\rm syst}).
\label{Ab_6}
\end{equation}
This measurement is obtained from data corresponding to 6.1 fb$^{-1}$ of
integrated luminosity collected with the D0 detector at
the Fermilab Tevatron collider. It is consistent with our previous measurement
\cite{D01} obtained with 1~fb$^{-1}$ and supersedes it.
The asymmetry disagrees with the prediction of the
SM by 3.2 standard deviations.
This is the first evidence for anomalous $CP$ violation in the mixing of neutral
$B$ mesons.

\begin{acknowledgments}
%
We thank the staffs at Fermilab and collaborating institutions,
and acknowledge support from the
DOE and NSF (USA);
CEA and CNRS/IN2P3 (France);
FASI, Rosatom and RFBR (Russia);
CNPq, FAPERJ, FAPESP and FUNDUNESP (Brazil);
DAE and DST (India);
Colciencias (Colombia);
CONACyT (Mexico);
KRF and KOSEF (Korea);
CONICET and UBACyT (Argentina);
FOM (The Netherlands);
STFC and the Royal Society (United Kingdom);
MSMT and GACR (Czech Republic);
CRC Program and NSERC (Canada);
BMBF and DFG (Germany);
SFI (Ireland);
The Swedish Research Council (Sweden);
and
CAS and CNSF (China).

\end{acknowledgments}


\begin{thebibliography}{99}

\bibitem{Nierste} A.~Lenz and U.~Nierste, J.~High Energy Phys.~{\bf0706}, 072 (2007).

\bibitem{pdg} C.~Amsler {\sl et al.}, Phys.~Lett.~B~{\bf 667}, 1 (2008), and
2009 partial update for the 2010 edition, and references therein.

\bibitem{sakharov} A.D.~Sakharov, Pisma Zh.~Eksp.~Teor.~Fiz. {\bf 5}, 32 (1967)
[Sov.~Phys.~JETP Lett. {\bf 5}, 24 (1967)].

\bibitem{gavela_huet} M.B.~Gavela, P.~Hernandez, J.~Orloff, O.~Pene, Mod. Phys.
Lett.~A {\bf 9}, 795 (1994); M.B.~Gavela, P.~Hernandez, J.~Orloff, O.~Pene, C.~Quimbay, 
Nucl. Phys.~B {\bf 430}, 382 (1994); P.~Huet and E.~Sather, Phys. Rev.~D {\bf 51}, 379 (1995).

\bibitem{d0asym2mu} V.A.~Abazov {\sl et al.} (D0 Collaboration),
arXiv:1005.2757 [hep-ex] (2010), submitted for publication in
Phys. Rev.~D.

\bibitem{d0det} V.M.~Abazov {\sl et al.} (D0 Collaboration),
Nucl. Instrum. Methods Phys. Res. A~{\bf 565}, 463  (2006).

\bibitem{Randall} L.~Randall and S.~Su, Nucl.~Phys.~B~\textbf{540}, 37 (1999).

\bibitem{Hewett} J.L.~Hewett, arXiv:hep-ph/9803370 (1998).

\bibitem{Hou} G.W.S.~Hou, arXiv:0810.3396 [hep-ph] (2008).

\bibitem{Soni} A.~Soni {\it et al.}, Phys. Lett. B {\bf 683}, 302 (2010); 
A.~Soni {\it et al.}, arXiv:1002.0595 (2010) [hep-ph] and references therein.

\bibitem{Buras} M.~Blanke, A.~J.~Buras, A.~Poschenrieder, C.~Tarantino, S.~Uhlig and A.~Weiler,
JHEP {\bf 0612}, 003 (2006).  W.~Altmannshofer, A.~J.~Buras, S.~Gori, P.~Paradisi and D.~M.~Straub, 
Nucl.\ Phys.\  B {\bf 830}, 17 (2010).

\bibitem{Grossman} Y.~Grossman {\it et al.}, Phys.~Rev.~Lett.~{\bf 97}, 151801 (2006).

\bibitem{rapidity} The D0 detector utilizes a right-handed coordinate system
with the $z$ axis pointing in the direction of the proton
beam and the $y$ axis pointing upwards. The azimuthal
angle  is defined in the $xy$ plane measured from the $x$
axis. The pseudorapidity is defined as $\eta \equiv 
-\ln{ \left[ \tan(\theta/2) \right] }$, where
$\theta$ is the polar angle with respect to the proton beam direction.

\bibitem{D01} V.M.~Abazov, \textit{et al.} (D0 Collaboration), Phys.~Rev.~D~\textbf{74}, 092001 (2006).

\bibitem{babar} B.~Aubert \textit{et al.} (Babar Collaboration), Phys.~Rev.~Lett.~\textbf{96}, 251802 (2006); 
B.~Aubert \textit{ et al.} (Babar Collaboration), arXiv:hep-ex/0607091 (2006).

\bibitem{belle} E.~Nakano \textit{et al.} (Belle Collaboration), Phys.~Rev.~D~\textbf{73}, 112002 (2006).

\bibitem{hfag} E.~Barberio {\it et al.} (HFAG), arXiv:0808.1297 [hep-ex] (2008).

\bibitem{asl-d0} V.M.~Abazov \textit{et al.} (D0 Collaboration), arXiv:0904.3907 [hep-ex], accepted for
publication in Phys.~Rev.~D.

\end{thebibliography}
\end{document}